\documentclass[epj]{svjourbh}
\usepackage{graphicx,amsmath,txfonts,cite}
\begin{document}

\title{Hydrogen and deuterium in shock wave experiments, ab initio simulations
and chemical picture modeling}
\author{Bastian Holst\inst{1} \and Ronald Redmer\inst{1} \and Victor K.
Gryaznov\inst{2,3} \and Vladimir E. Fortov\inst{2,3} 
    \and Igor L. Iosilevskiy\inst{2,4}}

\institute{  Universit\"at Rostock, Institut f\"ur Physik, 
          D-18051 Rostock, Germany 
\and 
Joint Institute for High Temperatures, Russian Academy of
Sciences, 125412 Moscow, Russia
\and
Institute of Problems of Chemical Physics, Russian Academy of
Sciences, 142432 Chernogolovka, Moscow Region, Russia
\and
Moscow Institute of Physics and Technology (State University), 
Joint Institute for High Temperatures, 141700 Dolgoprudny, Moscow Region, Russia
%\and
%Joint Institute for High Temperatures, Russian Academy of
%Sciences, 101600 Moscow, Russia
}

\mail{bastian.holst@uni-rostock.de}
\date{\today}

\abstract{
We present equation of state data of shock compressed hydrogen and deuterium.
These have been calculated in the physical picture by using {\it ab initio} 
molecular dynamics simulations based on finite temperature density functional 
theory as well as in the chemical picture via the Saha-D model. The results are
compared in detail with data of shock wave experiments obtained for condensed
and gaseous precompressed hydrogen and deuterium targets in a wide range of
shock compressions from low pressures up to megabars. 
}
% check!
% \pacs{31.15.Ar, 61.20.Ja, 62.50.+p, 64.30.+t, 72.20.-i, 71.30.+h} 
%\vspace*{5mm}
\maketitle

%%%%%%%%%%%%%%%%%%%%%
\section{Introduction}

The equation of state (EOS) of hydrogen and its isotopes has been in the focus
of research for many years for several reasons. In models of stellar and
planetary interiors~\cite{Guillot99, Guillot99a, Fortney10} hydrogen is the most
abundant element and its EOS is the most important component for the results.
Deuterium and tritium are target materials in inertial confinement fusion
experiments~\cite{Lindl04}. Therefore, a lot of experimental and theoretical
efforts were done to understand the behavior of hydrogen, deuterium, and tritium
in a wide range of densities and temperatures. Recent developments in shock-wave
experiments have enabled an access to a precise database in the megabar pressure
range. Single or multiple shock-wave experiments have been performed for
hydrogen (or deuterium) by using, e.g., high explosives~\cite{Mintsev06}, gas
guns~\cite{Nellis06}, pulsed power~\cite{Knudson01, Knudson03, Knudson04}, or
high-power lasers~\cite{Cauble97,Hicks09}. The strongly correlated states of
{\it warm dense matter} cover a wide range, from an atomic-molecular mixture at
low temperatures to fully ionized weakly coupled plasma at high temperatures. In
particular, the characterization of the transition region (partially ionized
plasma) is a great challenge both to theory and experiment since the bound
states exhibit a highly transient nature. This region of the phase diagram is,
however, of great relevance for planetary interiors. Important experimental
information is also gained from helioseismology data~\cite{Christensen02} that
allows to check and correct theoretical models very accurately in the high
temperature limit.

%A number of theoretical approaches has been developed that 
Some theoretical methods yield accurate results for limiting cases which then
can be used to benchmark more general but approximate methods. For instance, an
exact asymptotic expansion of thermodynamic functions can be given in the limit
of almost fully ionized, low density plasma~\cite{Larkin60, Starostin05}. {\it
Ab initio} simulation techniques such as Path Integral Monte Carlo
(PIMC)~\cite{Ceperley01, Delaney06}, quantum monte carlo
(QMC)~\cite{Lin09,Morales10, Morales10a} or finite-temperature density
functional theory molecular dynamics (FT-DFT-MD) simulations~\cite{Collins95,
Collins00} which treat quantum effects and correlations systematically have
taken a great benefit from the rapid progress in computing power. These methods
provide very accurate and reliable results for a variety of problems and
systems, especially for warm dense matter. In addition to these approaches,
advanced chemical models developed for partially ionized
plasmas~\cite{Ebeling69} have also been applied for warm dense matter for a long
time~\cite{Ebeling82, Saumon91, Saumon92, Saumon95, Reinholz95, Beule99,
Juranek00, Fortov03, Holst07, Gryaznov09}. 

In the present work we compare the results of the Saha-D model and of FT-DFT-MD 
simulations with shock-wave experiments for hydrogen and deuterium which were 
performed for different initial densities in a wide range of shock compressions. 
We find good agreement so that these models can also be used to give detailed 
predictions for high-pressure states that will be probed in future experiments 
by varying the initial conditions accordingly.  

Our paper is organized as follows. The FT-DFT-MD simulations are explained in 
Section~\ref{sec:MD} and the Saha-D model in Section~\ref{sec:Saha}. We present 
our results for the Hugoniot curves in Section~\ref{sec:results} and, finally, 
give a short summary in Section~\ref{sec:concl}.

%%%%%%%%%%%%%%%%%%%%%%%%%%%%%%%%%%%%%%%%%%%%%
\section{FT-DFT-MD simulations}\label{sec:MD}

FT-DFT-MD simulations are a powerful tool to describe warm dense
matter~\cite{Lenosky97, Collins01, Desjarlais02, Desjarlais03, Mazevet03,
Bonev04, Laudernet04, Mazevet05, Mazevet07, Vorberger07, Lorenzen09, Lorenzen10,
Caillabet11, Holst11}. Correlation and quantum effects are considered by a
combination of classical molecular dynamics for the ions and density functional
theory for the electrons. We use the plane wave density functional code VASP
(Vienna Ab Initio Simulation Package)~\cite{Kresse93, Kresse94, Kresse96} to
perform molecular dynamics simulations. VASP applies Mermin's finite temperature
density functional theory~\cite{Mermin65} which allows us to treat the electrons
even at higher temperatures on a quantum level. Projector augmented wave
potentials~\cite{Kresse99} were used and we applied a generalized gradient
approximation (GGA) within the parameterization of PBE~\cite{Perdew96}. The
plane wave cutoff $E_\text{cut}$ has to be chosen high enough to obtain
converged EOS data~\cite{Desjarlais03, Holst08}. A convergence of better than
1\% is secured for $E_\text{cut}=1200$~eV which was used in all calculations
presented here. In the MD scheme of VASP the Born-Oppenheimer approximation is
used, i.e. the dynamics of the ions is treated within a classical MD with
inter-ionic forces obtained by FT-DFT calculations via the Hellmann-Feynman
theorem. The electronic structure calculations were performed for a static array
of ions at each MD step. This was repeated until the EOS measures were converged
and a thermodynamic equilibrium was reached.

The simulations were done for 256~atoms in a supercell with periodic boundary
conditions. A Nos\'{e} thermostat~\cite{Nos'e84} controlled the temperature of
the ions, and the temperature of the electrons was fixed by Fermi weighting the
occupation of the electronic states~\cite{Kresse94}. Sampling of the Brillouin
zone using up to 14~\textbf{k}-points showed that well converged results were
obtained using Baldereschi's mean value point~\cite{Baldereschi73} for 256
particles. The same convergence behavior has previously been reported for
water~\cite{French09}. The size of the simulated supercell fixed the density of
the system. The internal energy was corrected due to zero point vibrations of
the molecules at low temperatures, taking into account quantum contributions of
a harmonic oscillator for each molecule~\cite{French09a}. For this procedure the
number of molecules has to be known, which was obtained by evaluating the pair
correlation function, see Ref.~\cite{Holst08} for details. The system was
simulated $1000-1500$~steps further after reaching the thermodynamic equilibrium
to ensure a small statistical error. The EOS data were then obtained by
averaging over all particles and simulation steps in equilibrium.

%%%%%%%%%%%%%%%%%%%%%%%%%%%%%%%%%%%%%%
\section{Saha-D model}\label{sec:Saha}

The Saha-D model EOS is based on the chemical picture~\cite{Ebeling69, Fortov03,
Mono80} which represents the plasma as a mixture of interacting electrons, ions,
atoms, and molecules. We consider the following components for hydrogen and
deuterium: $e^-$, $A$, $A^+$, $A_2$, $A_2^+$, ($A: H, D$). For this case the
Helmholtz free energy reads:
\begin{equation} \label{F=}
F( \{N_j\}, V, T) = \sum_j F_j^{(id)} + F_e^{(id)} + \Delta F_{C}^{(int)} +
\Delta F_{n}^{(int)} .
\end{equation}
We shortly outline the approximations in which these three contributions were
treated; for details, see~\cite{Fortov03, Gryaznov06}. The first two terms of
Eq.~(\ref{F=}) are the ideal gas contributions of heavy particles (atoms, ions,
and molecules) and of electrons. The latter corresponds to the partially
degenerate ideal Fermi gas~\cite{Landau76}. The last two terms describe
corrections due to Coulomb interactions and short-range interactions between
heavy particles. The Coulomb interaction effects of charged particles are
considered here within a modified pseudopotential approach \cite{Fortov03,
ILIEncy2004, Ilios80}. 

The electron-electron and ion-ion interaction is each treated by using the
Coulomb potential. For the effective electron-ion interaction we apply a
pseudopotential using the Glauberman form~\cite{Glaub1951}.  The parameters of
the pseudopotentials and of the corresponding pair correlation functions were
determined from the general conditions of local electroneutrality and dipole
screening, from the non-negativity constraint for the pair correlation
functions, and from a relation between the screening cloud amplitude and the
depth of the electron-ion pseudopotential. In the weak coupling limit, this
approximation coincides with the Debye model but in the strong coupling limit it
is much softer and demonstrates a quasi-crystalline behavior. 

At high densities as typical for shock-compressed hydrogen the short-range
repulsion between composite heavy particles ($A$, $A_2$, $A_2^+$) becomes very
important. This effect is taken into account in the Saha-D model within a simple
soft-sphere approximation~\cite{Young77} which is modified for a mixture of soft
spheres with different radii. In this case the effective packing fraction $Y$ is
calculated via the individual diameters $\sigma_j$ of each particle species in
correspondence with the one-fluid approximation:
\begin{eqnarray} \label{Y=}
Y &=& \frac{\pi n \sigma_c^3}{6} ,\quad 
\sigma_c = \left( \frac{\sum_j n_j \sigma_j^3}{n} \right)^{1/3} ,\quad 
n=\sum_j n_j \,.
\end{eqnarray}
$\sigma_j$ is the diameter of the soft spheres in the respective potential
$V_{SS}(r) = \epsilon (r/\sigma_j)^{-s}$. The contribution of the intermolecular
repulsion dominates the EOS of dense hydrogen and deuterium in a wide range of
pressures at low temperatures. The contribution of atom-atom and atom-molecule
repulsion becomes important at elevated temperatures. The parameters for the
soft-sphere repulsion for $A_2 - A_2$, $A_2 - A$, and $A - A$ are chosen
according to the spherically symmetric parts of the effective interaction
potentials of the non-empirical atom-atom approximation~\cite{Yakub99}. The key
parameter of this approximation is the ratio of corresponding soft-sphere
diameters for atoms and molecules, $\sigma_A/\sigma_{A_2}$. This ratio
determines the change of intrinsic volumes of two atoms in comparison with that
of a molecule ($2V_A/V_{A_2}$); it determines the effective shift of the
dissociation and ionization equilibrium in warm dense hydrogen and deuterium.
All parameters of the soft-sphere repulsion are given in Table~\ref{tab:1}. The
parameter $\epsilon$ is chosen such that our soft-sphere potential for
molecule-molecule repulsion will be close to the potential~\cite{Yakub99} at a
distance $r=2a_0$ (in this case $\epsilon=0.138eV$).

\begin{table}
\caption{\label{tab:1} Parameters of $A_2 - A_2$, $A_2^+ - A_2^+$, $A - A$
repulsion in Eq.~(\ref{Y=}); $a_0$ is the Bohr radius.}
\begin{center}
\begin{tabular}{ccccc}
\hline\noalign{\smallskip}
 &$s$&$\sigma_{j}/a_0$\\
\noalign{\smallskip}\hline\noalign{\smallskip}
  $A_2$ & 6 & 4.0 \\
  $A$   & 6 & 3.2 \\
$A_2^+$ & 6 & 3.2 \\
\noalign{\smallskip}\hline
 \end{tabular}
\end{center}
% \end{ruledtabular}
\end{table}

In order to take into account the existence of condensed states (liquid and
solid) for hydrogen and deuterium, the attraction term in the free energy has to
be considered together with the soft-sphere repulsion. In our case $\Delta
F_n^{(int)}$ reads:
\begin{eqnarray} \label{Fcold=}
\Delta F_n^{(int)} &=& \Delta F_{ss} + \Delta F_{attr} ,\\
\Delta F_{attr} &=& - BN_{molecules}^{(1+\delta)}V^{-\delta} .\nonumber
% \quad B,\delta = const \mathrm{.}
\end{eqnarray}
The attractive corrections~\cite{Gryaznov98} are independent of temperature. The
choice of $\delta=1$ as in our case corresponds to a van der Waals-like
approximation. The parameter $B$ supplies the correct sublimation energy of a
molecular system in the condensed state.
% B=??? RR

%%%%%%%%%%%%%%%%%%%%%%%%%%%%%%%%%%%%
\section{Results}\label{sec:results}

\begin{figure}[htb]
\centering
\resizebox{0.75\columnwidth}{!}{
\includegraphics[width=1.0\columnwidth]{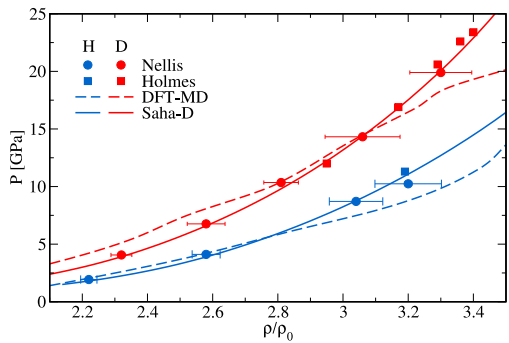}}
\caption{Shock compression of liquid hydrogen (blue) and deuterium (red). 
The Hugoniot curves obtained with the Saha-D model (solid line) and the 
FT-DFT-MD (dashed line) are compared with shock wave experiments of 
Nellis {\it et al.}~\cite{Nellis83} (circles) and 
Holmes {\it et al.}~\cite{Holmes95} (squares)}.
\label{fig:gasgunp}
\end{figure}

We calculated Hugoniot curves based on the FT-DFT-MD and Saha-D EOS data
sets for hydrogen and deuterium. In Fig.~\ref{fig:gasgunp} we compare our 
calculations with gas-gun experiments~\cite{Nellis83,Holmes95} on liquid 
hydrogen and deuterium. The experiments for deuterium and hydrogen are 
reproduced by the Saha-D model within the uncertainties of the experiments. 
The FT-DFT-MD Hugoniot curves reproduce the experiments with less precision. 
In the case of hydrogen the compression rate is slightly overestimated above 
5~GPa. 
% ok? RR 
% at the experimental results at about 10 GPa. 
For deuterium this occurs for pressures above 15~GPa. Furthermore, the 
FT-DFT-MD curve lies above the experiments for pressures below 10~GPa. 
Between 10~GPa and 20~GPa the experimental data are reproduced within
error bars. 

The less precision of the FT-DFT-MD data in the region of the gas gun
experiments is connected with the relatively abrupt onset of dissociation
processes which lead to an increase of the compression ratio at about 20~GPa and
4200~K as it has already been reported~\cite{Tamblyn10}. The reason for this
behavior can be related to the underestimation of the fundamental band gap in
DFT-based electronic structure calculations. More accurate exchange-correlation
functionals than PBE, specifically derived for finite temperatures, are urgently
needed for warm dense matter. QMC calculations treat the exchange-correlation
directly and are not affected by this approximation. Recent QMC calculations
find in fact a shift of the dissociation region compared to
DFT~\cite{Morales10a} but show no different results for the EOS for conditions
relevant for planetary interiors~\cite{Morales10, Morales12}.   For instance,
the Saha-D model yields 10\% dissociation at 60~GPa and a temperature of 13000~K
along the deuterium Hugoniot curve. Along the hydrogen Hugoniot curve
dissociation occurs at about 15~GPa and 3000~K within the FT-DFT-MD model. The
Saha-D model predicts 10\% dissociation at 50~GPa and a temperature of 14000~K
along the hydrogen Hugoniot curve.

The presented theoretical predictions both agree with the experiments within
10\% accuracy in compression ratio and can therefore describe the principal
behavior of the obtained results well. The compression reached in shock
compressed hydrogen is higher than in deuterium at the same pressure. This is
reproduced by the calculated Hugoniot curves, see also~\cite{Caillabet11}. 

The FT-DFT-MD hydrogen EOS data were also used to calculate the deuterium
Hugoniot curve. To adjust for the deuterium initial conditions we considered the
initial conditions of the hydrogen EOS at half the deuterium density given in
the experiment ($0.171$~g/cm$^3$), i.e.\ $0.0855$~g/cm$^3$. Plotting the
resulting pressure versus the compression ratio, both Hugoniot curves are almost
the same. On the other hand, calculating the pressure with the deuterium EOS and
adjusting for the hydrogen initial conditions in the same way, the resulting
Hugoniot curves are identical within a smaller error than the statistical error
of the FT-DFT-MD simulations. The different compression ratios of hydrogen and
deuterium as seen in the experiment are, therefore, only slightly caused by
differences in the EOS data of both isotopes at warm dense matter conditions.
The difference in the compression ratios is mainly due to the fact that the
densities of the liquid targets at $20$~K do not scale exactly by a factor of
two. Scaling the deuterium density to that of hydrogen the initial density of
liquid deuterium would be $0.0855$~g/cm$^3$ which differs from the value
relevant for liquid hydrogen which is $0.071$~g/cm$^3$. This deviation of about
20\% entails the different Hugoniot curves. 

\begin{figure}[htb]
\centering
\resizebox{0.75\columnwidth}{!}{
\includegraphics[width=1.0\columnwidth]{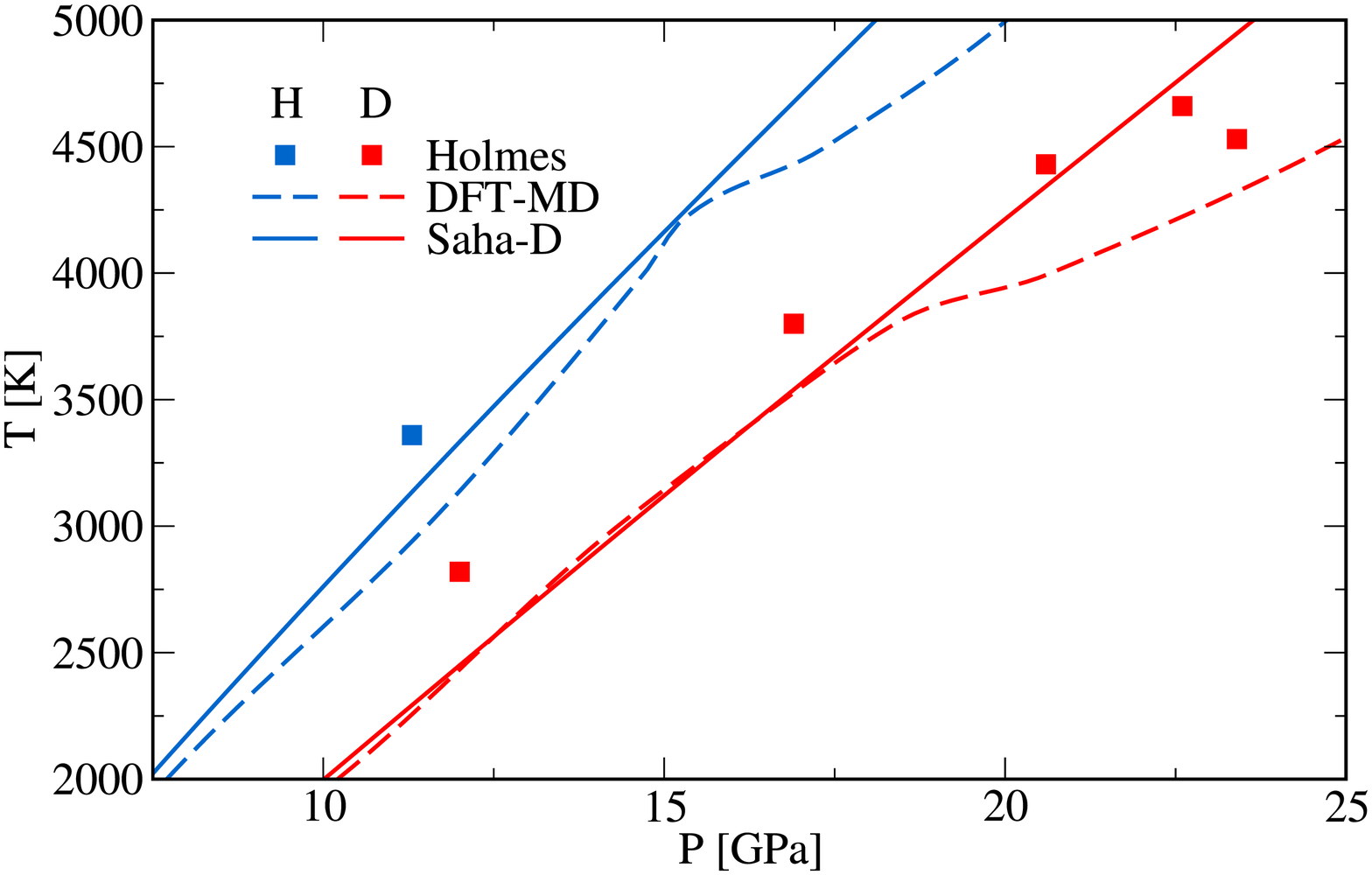}}
\caption{Shock compression of liquid hydrogen (blue) and deuterium (red). 
Temperatures along the Hugoniot curves as obtained with the Saha-D model 
(solid line) and FT-DTF-MD (dashed line) are displayed as function of 
pressure and compared with shock-wave experiments (squares)~\cite{Holmes95}. }
\label{fig:gasgunpt}
\end{figure}

The temperature along the Hugoniot curves as predicted by the theoretical models
is shown in Fig.~\ref{fig:gasgunpt} and compared with gas-gun experimental
data~\cite{Holmes95}. The temperatures measured in the experiments are in
general higher than predicted by the Saha-D model and the FT-DFT-MD simulations,
except for two deuterium points above 22~GPa which are below the Saha-D curve.
Again, onset of dissociation causes the slight kink in the FT-DFT-MD curves at
about 18~GPa (D$_2$) and 15~GPa (H$_2$). The general behavior indicated by the
experiments can be reproduced: the temperature along the hydrogen Hugoniot curve
is higher than that for deuterium at the same pressure. The maximum deviation of
both theoretical models from the experimental data is about $400$~K. 

\begin{figure}[htb]
\centering
\resizebox{0.75\columnwidth}{!}{
\includegraphics[width=1.0\columnwidth]{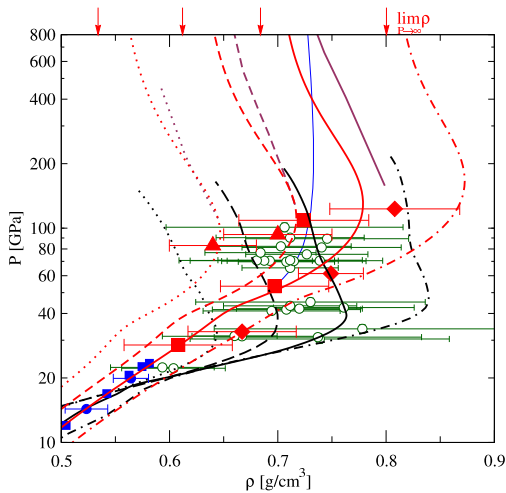}}
\caption{Deuterium single shock principal Hugoniot curves starting from
different 
initial densities (gaseous, liquid, and solid) as derived from FT-DFT-MD
(black),
Saha-D~\cite{Gryaznov09} (red), RPIMC~\cite{Militzer00} (blue), 
DPIMC~\cite{Levashov07} (dark red). Dotted line: $\rho_0=0.1335$~g/cm$^3$, 
dashed line: $\rho_0=0.153$~g/cm$^3$, solid line: $\rho_0=0.171$~g/cm$^3$, 
dot-dashed line: $\rho_0=0.199$~g/cm$^3$. Shock wave experiments: 
Nellis {\it et al.}~\cite{Nellis83} (blue circles), 
Holmes {\it et al.}~\cite{Holmes95} (blue squares), 
Knudson {\it et al.}~\cite{Knudson01,Knudson03} (open green circles), 
Grishechkin {\it et al.}~\cite{Grish04} (red triangles),
Boriskov {\it et al.}~\cite{Boriskov05} (red squares and diamonds). 
The arrows at the top show the limiting compression for ultra-high pressures
($4\times\rho_0$) for each principal Hugoniot.}
\label{fig:dhugprho}
\end{figure}

We have also applied both theoretical EOS data sets to calculate the Hugoniot
curves for different initial conditions in order to study the compression
behavior of the hydrogen isotopes for a wide range of densities off the
principal Hugoniot. Fig.~\ref{fig:dhugprho} shows Hugoniot curves with respect
to initial conditions as chosen in recent shock-wave experiments with
precompressed targets~\cite{Grish04,Boriskov05}. Two experiments were performed
with gaseous targets at 1.5~kbar ($\rho_0=0.1335$~g/cm$^3$) and 2.0~kbar
($\rho_0=0.153$~g/cm$^3$). Two other data sets were obtained with liquid
($\rho_0=0.171$~g/cm$^3$) and solid ($\rho_0=0.199$~g/cm$^3$) deuterium targets.

The two theoretical results and the experiments show the same general behavior:
the attained absolute density is higher the more precompressed the target is. It
has to be pointed out that the maximum compression ratio shows the inverse
behavior, it decreases with higher precompression. Even so, the maximum density
that can be probed in single shock experiments increases with precompression.
The theoretical predictions of the two methods for the maximum density agree at
the conditions of the experiments with gaseous and liquid targets and range from
0.65~g/cm$^3$ to 0.775~g/cm$^3$. For the initial condition in the solid, there
is a slight difference: FT-DFT-MD predicts 0.83~g/cm$^3$ and Saha-D
0.87~g/cm$^3$. The pressure at this maximum compression density is also
different for the two models; it ranges from 30 to 50~GPa within FT-DFT-MD and
from 80 to 150~GPa according to Saha-D. These values cannot be discriminated via
the few experimental points. 

\begin{figure}[htb]
\centering
\resizebox{0.75\columnwidth}{!}{
\includegraphics[width=1.0\columnwidth]{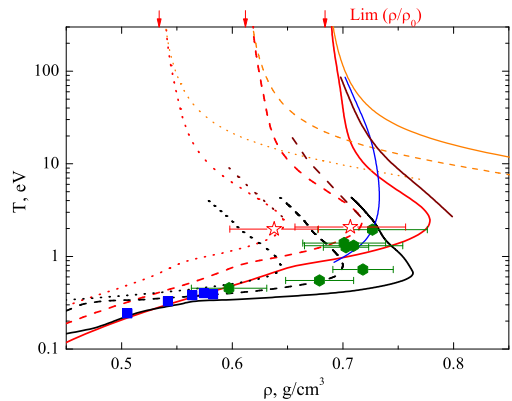}}
\caption{Temperature of shock compressed deuterium for different initial
conditions predicted by FT-DFT-MD (black), Saha-D~\cite{Gryaznov09} (red), 
RPIMC~\cite{Militzer00} (blue), DPIMC~\cite{Levashov07} (dark red), and the
asymptotically strict Saha-S~\cite{Gryaznov06} (orange) model in comparison 
with experiments using liquid [Holmes {\it et al.}~\cite{Holmes95} 
(blue squares) and Bailey {\it et al.}~\cite{Bailey08} (green hexagons)]  
and gaseous targets [Grishechkin {\it et al.}~\cite{Grish04} (red stars)]. 
Dotted, dashed, and solid lines correspond to initial deuterium densities 
of $\rho_0=0.1335, 0.153, 0.171$~g/cm$^3$, respectively. Arrows indicate 
the limiting compression for ultra-high pressures for each Hugoniot curve.}
\label{fig:dhugtrho}
\end{figure}

Fig.~\ref{fig:dhugtrho} shows the temperatures along the Hugoniot curves of
deuterium using the initial conditions of the experiments with liquid and 
precompressed gaseous targets~\cite{Holmes95, Grish04, Bailey08}.

We have to note again that with a higher precompression also a higher density
can be reached. The measurements indicate a temperature of about 2~eV for all
three initial conditions. These results can be reproduced by the theoretical
models within the error bars. The densest states are reached for the liquid
deuterium target. The experiments of Bailey {\it et al.}~\cite{Bailey08} show a
maximum density between 0.7 and 0.8~g/cm$^3$ which is reproduced by both
theories. The temperatures measured in the experiments are underestimated by
FT-DFT-MD and overestimated by Saha-D, with increasing deviation for higher
temperatures. The limit of the Saha-D model at high temperatures can be checked
by comparing with results obtained by the Saha-S model~\cite{Gryaznov06} which
is asymptotically exact for $\Gamma_D \ll 1, n\lambdabar_e^3 \ll 1$. Such a
comparison shows that Saha-D (together with PIMC~\cite{Militzer00,Levashov07})
yields the correct high-temperature limit. The deviation of the Saha-S from the
Saha-D curves at lower temperatures is due to the fact that the Saha-S model is
no longer applicable for these parameters. In particular, the Saha-S model does
not take into account the short-range repulsion effects of composite particles
($A, A_2, A_2^+$). Interestingly, the temperature at maximum compression is
almost independent of the initial conditions. The predictions of the theoretical
models show that the curves are shifted only to higher densities while the
temperature remains almost constant. This finding is supported by the
experimental results.

%%%%%%%%%%%%%%%%%%%%%
\section{Conclusions}\label{sec:concl}

We have calculated the EOS of deuterium and hydrogen with FT-DFT-MD simulations
in the physical picture and the Saha-D model in the chemical picture over a wide
range of density and temperature which enabled us to compare those results with
recent shock-wave experiments. The theories predict, in agreement with the
experiments, that higher compressed densities can be reached using precompressed
targets, while the maximum compression ratio decreases. We compare also the
temperature along the Hugoniot curve of deuterium with experimental data and
find that only the density is affected by precompression, while the temperature
remains almost the same along the different pathways. This leads to an increased
pressure with higher precompression along the Hugoniot.

% Outlook experiments?
% Probe off-Hugoniot points?
A check of EOS models against experiments within the WDM regime is available
mostly for the relatively limited density and temperature range along the
principal Hugoniot curve. Experiments producing shock waves within precompressed
targets enable to check the quality of EOS models in a wider range in the phase
diagram. Both EOS (based on FT-DFT-MD simulations and the Saha-D model) could
reproduce the experimental data. On the other hand, neither the Saha-D model,
which uses effective two-particle potentials with parameters that have been
chosen to match physical constraints, nor the FT-DFT-MD method, which has no
adjustable parameters, can reproduce all experimental features precisely.
Nevertheless, experimental data is still not available in the needed quantity
and precision to allow for a definite decision of which model has to be used to
describe all quantities in all ranges of the phase diagram. We therefore look
forward to future high-pressure experiments, especially off the principal
Hugoniot curve. 

% Relevant for what???
The combination of an advanced chemical model with an {\it ab initio} approach
yields a reasonable description of warm dense hydrogen because the low-density
molecular liquid, the strongly correlated warm dense fluid, as well as the hot
plasma can be described adequately, see also~\cite{Redmer10}. Simultaneously
this combination saves computational power as the treatment of a low-density
molecular liquid is increasingly demanding when using FT-DFT-MD simulations. The
treatment of a free energy minimiztion model like Saha-D is much less expensive
regarding computational time. Accordingly, this combination provides an
opportunity to construct a wide-range EOS for planetary interior modelling for
which a database from ambient conditions up to pressures of several tens of
megabar and temperatures up to about 100.000~K is needed. This project is going
to be compiled for and will be be applied to model the interior of
Jupiter~\cite{Nettelmann12}.

\begin{acknowledgement}
This work was supported by the Deutsche Forschungsgemeinschaft within
the SFB~652, the High Performance Computing Center North (HLRN), 
and the Program of the Presidium of the Russian Academy of Sciences 
''Research of Matter at Extreme Conditions''. We acknowledge support
from the computer center of the University of Rostock and of the 
Education Center -- Physics of High Energy Density Matter -- of the 
Moscow Institute of Physics and Technology. We thank Eugene Yakub for 
helpful discussions and for providing us with the results of calculations 
for the deuterium Hugoniot. 
% further people thanks?
\end{acknowledgement}

%%%%%%%%%%%%%%%%%%
\bibliographystyle{epj}
\bibliography{qmd_sh}

\begin{thebibliography}{77}

\bibitem{Guillot99}
T.~Guillot, Science \textbf{286}, 72 (1999)

\bibitem{Guillot99a}
T.~Guillot, Planet. Space Sci. \textbf{47}, 1183 (1999)

\bibitem{Fortney10}
J.J. Fortney, N.~Nettelmann, Space Sci. Rev. \textbf{152}, 423 (2010)

\bibitem{Lindl04}
J.D. Lindl, P.~Amendt, R.L. Berger, S.G. Glendinning, S.H. Glenzer, S.W. Haan,
  R.L. Kauffman, O.L. Landen, L.J. Suter, Phys. Plasmas \textbf{11}, 339 (2004)

\bibitem{Mintsev06}
V.B. Mintsev, V.E. Fortov, J. Phys. A: Math. Gen. \textbf{39}, 4319 (2006)

\bibitem{Nellis06}
W.J. Nellis, Rep. Prog. Phys. \textbf{69}, 1479 (2006)

\bibitem{Knudson01}
M.D. Knudson, D.L. Hanson, J.E. Bailey, C.A. Hall, J.R. Asay, W.W. Anderson,
  Phys. Rev. Lett. \textbf{87}, 225501 (2001)

\bibitem{Knudson03}
M.D. Knudson, D.L. Hanson, J.E. Bailey, C.A. Hall, J.R. Asay, Phys. Rev. Lett.
  \textbf{90}, 035505 (2003)

\bibitem{Knudson04}
M.D. Knudson, D.L. Hanson, J.E. Bailey, C.A. Hall, J.R. Asay, C.~Deeney, Phys.
  Rev. B \textbf{69}, 144209 (2004)

\bibitem{Cauble97}
R.~Cauble, L.B.D. Silva, T.S. Perry, D.R. Bach, K.S. Budil, P.~Celliers, G.W.
  Collins, A.~Ng, T.W. {Barbee Jr.}, B.A. Hammel et~al., Phys. Plasmas
  \textbf{4}, 1857 (1997)

\bibitem{Hicks09}
D.G. Hicks, T.R. Boehly, P.M. Celliers, J.H. Eggert, S.J. Moon, D.D.
  Meyerhofer, G.W. Collins, Phys. Rev. B \textbf{79}, 014112 (2009)

\bibitem{Christensen02}
J.~Christensen-Dalsgaard, Rev.~Mod.~Phys. \textbf{74}, 1073 (2002)

\bibitem{Larkin60}
A.~Larkin, Sov. Phys. JETP \textbf{11}, 1363 (1960)

\bibitem{Starostin05}
A.~Starostin, V.~Roerich, JETP \textbf{100}, 186 (2005)

\bibitem{Ceperley01}
D.M. Ceperley, E.~Manousakis, J. Chem. Phys. \textbf{115}, 10111 (2001)

\bibitem{Delaney06}
K.T. Delaney, C.~Pierleoni, D.M. Ceperley, Phys. Rev. Lett. \textbf{97}, 235702
  (2006)

\bibitem{Lin09}
F.~Lin, M.A. Morales, K.T. Delaney, C.~Pierleoni, R.M. Martin, D.M. Ceperley,
  Phys. Rev. Lett. \textbf{103}, 256401 (2009)

\bibitem{Morales10}
M.A. Morales, C.~Pierleoni, D.M. Ceperley, Phys. Rev. E \textbf{81}, 021202
  (2010)

\bibitem{Morales10a}
M.A. Morales, C.~Pierleoni, E.~Schwegler, D.M. Ceperley, Proc. Natl. Acad. Sci.
  U.S.A. \textbf{107}, 12799 (2010)

\bibitem{Collins95}
L.A. Collins, I.~Kwon, J.D. Kress, N.J. Troullier, D.~Lynch, Phys. Rev. E
  \textbf{52}, 6202 (1995)

\bibitem{Collins00}
L.A. Collins, J.D. Kress, S.R. Bickham, T.J. Lenosky, N.J. Troullier, High
  Pressure Research \textbf{16}, 313 (2000)

\bibitem{Ebeling69}
W.~Ebeling, Physica \textbf{43}, 293 (1969)

\bibitem{Ebeling82}
W.~Ebeling, W.~Richert, Annalen der Physik \textbf{39}, 362 (1982)

\bibitem{Saumon91}
D.~Saumon, G.~Chabrier, Phys. Rev. A \textbf{44}, 5122 (1991)

\bibitem{Saumon92}
D.~Saumon, G.~Chabrier, Phys. Rev. A \textbf{46}, 2084 (1992)

\bibitem{Saumon95}
D.~Saumon, G.~Chabrier, H.M. {van Horn}, Astrophys. J. Suppl. Ser. \textbf{99},
  713 (1995)

\bibitem{Reinholz95}
H.~Reinholz, R.~Redmer, S.~Nagel, Phys. Rev. E \textbf{52}, 5368 (1995)

\bibitem{Beule99}
D.~Beule, W.~Ebeling, A.~F\"orster, H.~Juranek, S.~Nagel, R.~Redmer,
  G.~R\"opke, Phys. Rev. B \textbf{59}, 14177 (1999)

\bibitem{Juranek00}
H.~Juranek, R.~Redmer, J. Chem. Phys. \textbf{112}, 3780 (2000)

\bibitem{Fortov03}
V.E. Fortov, V.Y. Ternovoi, M.V. Zhernokletov, M.A. Mochalov, A.L. Mikhailov,
  A.S. Filimonov, A.A. Pyalling, V.B. Mintsev, V.K. Gryaznov, I.L. Iosilevskii,
  J. Exp. Theor. Phys. \textbf{97}, 259 (2003)

\bibitem{Holst07}
B.~Holst, N.~Nettelmann, R.~Redmer, Contrib. Plasma Phys. \textbf{47}, 368
  (2007)

\bibitem{Gryaznov09}
V.K. Gryaznov, I.L. Iosilevskiy, J.~Phys.~A: Math. \& Gen. \textbf{42}, 214007
  (2009)

\bibitem{Lenosky97}
T.J. Lenosky, J.D. Kress, L.A. Collins, Phys. Rev. B \textbf{56}, 5164 (1997)

\bibitem{Collins01}
L.A. Collins, S.R. Bickham, J.D. Kress, S.~Mazevet, T.J. Lenosky, N.J.
  Troullier, W.~Windl, Phys. Rev. B \textbf{63}, 184110 (2001)

\bibitem{Desjarlais02}
M.P. Desjarlais, J.D. Kress, L.A. Collins, Phys. Rev. E \textbf{66}, 025401
  (2002)

\bibitem{Desjarlais03}
M.P. Desjarlais, Phys. Rev. B \textbf{68}, 064204 (2003)

\bibitem{Mazevet03}
S.~Mazevet, J.D. Kress, L.A. Collins, P.~Blottiau, Phys. Rev. B \textbf{67},
  054201 (2003)

\bibitem{Bonev04}
S.A. Bonev, B.~Militzer, G.~Galli, Phys.\ Rev.\ B \textbf{69}, 014101 (2004)

\bibitem{Laudernet04}
Y.~Laudernet, J.~Cl{\'e}rouin, S.~Mazevet, Phys. Rev. B \textbf{70}, 165108
  (2004)

\bibitem{Mazevet05}
S.~Mazevet, M.P. Desjarlais, L.A. Collins, J.D. Kress, N.H. Magee, Phys. Rev. E
  \textbf{71}, 016409 (2005)

\bibitem{Mazevet07}
S.~Mazevet, F.~Lambert, F.~Bottin, G.~Z{\'e}rah, J.~Cl{\'e}rouin, Phys. Rev. E
  \textbf{75}, 056404 (2007)

\bibitem{Vorberger07}
J.~Vorberger, I.~Tamblyn, B.~Militzer, S.A. Bonev, Phys. Rev. B \textbf{75},
  024206 (2007)

\bibitem{Lorenzen09}
W.~Lorenzen, B.~Holst, R.~Redmer, Phys. Rev. Lett. \textbf{102}, 115701 (2009)

\bibitem{Lorenzen10}
W.~Lorenzen, B.~Holst, R.~Redmer, Phys. Rev. B \textbf{82}, 195107 (2010)

\bibitem{Caillabet11}
L.~Caillabet, S.~Mazevet, P.~Loubeyre, Phys. Rev. B \textbf{83}, 094191 (2011)

\bibitem{Holst11}
B.~Holst, M.~French, R.~Redmer, Phys. Rev. B \textbf{83}, 235120 (2011)

\bibitem{Kresse93}
G.~Kresse, J.~Hafner, Phys.\ Rev.\ B \textbf{47}, 558 (1993)

\bibitem{Kresse94}
G.~Kresse, J.~Hafner, Phys.\ Rev.\ B \textbf{49}, 14251 (1994)

\bibitem{Kresse96}
G.~Kresse, J.~Furthm\"uller, Phys.\ Rev.\ B \textbf{54}, 11169 (1996)

\bibitem{Mermin65}
N.D. Mermin, Phys.\ Rev. \textbf{137}, A1441 (1965)

\bibitem{Kresse99}
G.~Kresse, D.~Joubert, Phys.\ Rev.\ B \textbf{59}, 1758 (1999)

\bibitem{Perdew96}
J.P. Perdew, K.~Burke, M.~Ernzerhof, Phys.\ Rev.\ Lett. \textbf{77}, 3865
  (1996)

\bibitem{Holst08}
B.~Holst, R.~Redmer, M.P. Desjarlais, Phys. Rev. B \textbf{77}, 184201 (2008)

\bibitem{Nos'e84}
S.~Nos\'{e}, J. Chem. Phys. \textbf{81}, 511 (1984)

\bibitem{Baldereschi73}
A.~Baldereschi, Phys. Rev. B \textbf{7}, 5212 (1973)

\bibitem{French09}
M.~French, T.R. Mattsson, N.~Nettelmann, R.~Redmer, Phys. Rev. B \textbf{79},
  054107 (2009)

\bibitem{French09a}
M.~French, R.~Redmer, Journal of Physics: Condensed Matter \textbf{21}, 375101
  (2009)

\bibitem{Mono80}
V.~Gryaznov, I.~Iosilevskiy, Y.~Krasnikov, N.~Kuznetsova, V.~Kucherenko,
  G.~Lappo, B.~Lomakin, G.~Pavlov, E.~Son, V.~Fortov, \emph{Thermophysics of
  Gas Core Nuclear Reactor} (Atomizdat, Moscow, 1980)

\bibitem{Gryaznov06}
V.K. Gryaznov, S.V. Ayukov, V.A. Baturin, I.L. Iosilevskiy, A.N. Starostin,
  V.E. Fortov, J.~Phys.~A: Math. \& Gen. \textbf{39}, 4459 (2006)

\bibitem{Landau76}
L.~Landau, E.~Lifshits, \emph{Statistical physics}, Vol.~5, 3rd~edn. (Nauka,
  Moscow, 1976)

\bibitem{ILIEncy2004}
I.~Iosilevskiy, in \emph{Encyclopedia of Low-Temperature Plasma Physics.
  Supplement}, edited by A.~Starostin, I.~Iosilevskiy (Fizmatlit, Moscow,
  2004), pp. 349--428

\bibitem{Ilios80}
I.L. Iosilevskii, High Temp \textbf{18}, 355 (1980)

\bibitem{Glaub1951}
A.~Glauberman, Dokl. Acad. Nauk SSSR \textbf{78}, 883 (1951)

\bibitem{Young77}
D.~Young, Tech. Rep. UCRL-52352, LLNL, Univ. California (1977)

\bibitem{Yakub99}
E.~S.Yakub, Physica B - Cond. Matter \textbf{265}, 31 (1999)

\bibitem{Gryaznov98}
V.K. Gryaznov, V.E. Fortov, M.V. Zhernokletov, G.V. Simakov, R.F. Trunin, L.I.
  Trusov, I.L. Iosilevskiy, J.~Exp.~Theor.~Phys \textbf{87}, 678 (1998)

\bibitem{Nellis83}
W.J. Nellis, A.C. Mitchell, M.~{van Thiel}, G.J. Devine, R.J. Trainor,
  N.~Brown, J. Chem. Phys. \textbf{79}, 1480 (1983)

\bibitem{Holmes95}
N.~Holmes, M.~Ross, W.~Nellis, Phys. Rev. B \textbf{52}, 15835 (1995)

\bibitem{Tamblyn10}
I.~Tamblyn, S.A. Bonev, Phys. Rev. Lett. \textbf{104}, 065702 (2010)

\bibitem{Morales12}
M.A. Morales, L.X. Benedict, D.S. Clark, E.~Schwegler, I.~Tamblyn, S.A. Bonev,
  A.A. Correa, S.W. Haan, High Energy Density Physics \textbf{8}, 5  (2012),
  ISSN 1574-1818

\bibitem{Grish04}
S.~Grishechkin, S.~Gruzdev, V.~Gryaznov, M.~Zhernokletov, R.~Il'kaev,
  I.~Iosilevskii, G.~Kashintseva, S.~Kirshanov, S.~Manachkin, V.~Mintsev
  et~al., JETP~Letters \textbf{80}, 398 (2004)

\bibitem{Boriskov05}
G.~Boriskov, A.~Bykov, R.~Il'kaev, V.~Selemir, G.V. Simakov, R.~Trunin,
  V.~Urlin, A.~Shuikin, W.~Nellis, Phys. Rev. B \textbf{71}, 092104 (2005)

\bibitem{Militzer00}
B.~Militzer, D.M. Ceperley, Phys. Rev. Lett. \textbf{85}, 1890 (2000)

\bibitem{Levashov07}
P.R. Levashov, V.S. Filinov, A.V. Botsan, M.~Bonitz, V.E. Fortov, \emph{First
  principal calculation of deuterium Hugoniots by Path integral Monte-Carlo
  method}, in \emph{IX Kharitons Topical Scientific Readings. Extreme States of
  Substance. Detonation. Shock Waves}, edited by A.L. Mikhailov (RFNC-VNIIEF,
  2007), pp. 276--281

\bibitem{Bailey08}
J.E. Bailey, M.D. Knudson, A.L. Carlson, G.S. Dunham, M.P. Desjarlais, D.L.
  Hanson, J.R. Asay, Phys. Rev. B \textbf{78}, 144107 (2008)

\bibitem{Redmer10}
R.~Redmer, G.~R\"opke, Contrib. Plasma Phys. \textbf{50}, 970 (2010)

\bibitem{Nettelmann12}
N.~Nettelmann, A.~Becker, B.~Holst, R.~Redmer, Astrophys. J. \textbf{750}, 52
  (2012)

\end{thebibliography}

\end{document}